\documentclass[aps,amsfonts,showpacs]{revtex4}
\usepackage{amsmath}
\usepackage{graphicx}

\begin{document}

\def\be{\begin{equation}}
\def\ee{\end{equation}}
\def\bea{\begin{eqnarray}}
\def\eea{\end{eqnarray}}
\def\bml{\begin{mathletters}}
\def\eml{\end{mathletters}}
\def\l{\label}
\def\b{\bullet}
\def\eqn#1{(~\ref{eq:#1}~)}
\def\no{\nonumber}
\def\av#1{{\langle  #1 \rangle}}
\def\m{{\rm{min}}}
\def\M{{\rm{max}}}

\title{Adaptation dynamics of the quasispecies model}

\author{Kavita Jain}
\affiliation{Theoretical Sciences Unit, Jawaharlal Nehru Centre for
Advanced Scientific Research, Jakkur P.O., Bangalore
560064,India \footnote{Also at Evolutionary and Organismal Biology
Unit}}

\widetext
\date{\today}

\begin{abstract}
We study the adaptation dynamics of an initially maladapted population
evolving via the elementary processes of mutation and selection. The
evolution occurs on rugged fitness landscapes which are defined on the
multi-dimensional genotypic space and have many local peaks separated
by low fitness valleys. We  mainly focus on the Eigen's model that
describes the deterministic dynamics of an infinite number of
self-replicating molecules. In the stationary state, for small mutation
rates such a population forms a {\it quasispecies} which consists of
the fittest genotype and its closely related mutants. The quasispecies
dynamics on rugged fitness landscape follow a punctuated (or
step-like) pattern in which a population jumps from a low fitness peak
to a higher one, stays there for a considerable time before shifting
the peak again and eventually reaches the global maximum of the
fitness landscape. We calculate exactly several properties of this
dynamical process within a simplified version of the quasispecies
model. 
\end{abstract}
\maketitle

\section{Introduction}

Consider a maladapted population such as a bacterial colony in a
glucose-limited environment, or a viral population in a vaccinated
animal cell. In such harsh environments, the less fit members of the
population are likely to perish and only the highly fit ones can
survive to the next 
generation. In this manner, the fitness of the population increases
with time and the initially maladapted population evolves to a
well-adapted state. In the last century, there has been a concerted effort to 
put this verbal theory of Darwin \cite{Darwin:1859} on a solid quantitative
footing by performing long-term experiments on microbial populations and
studying theoretical models of biological evolution. 

One of the questions in evolutionary biology concerns the mode of
evolution. In the experiments on microbes, it is found that the fitness
of the maladapted population can 
increase with time in either a smooth continuous manner
\cite{Novella:1995} or sudden
jumps \cite{Elena:2003a}. The latter mode  is consistent
with evolution on a fitness landscape defined on genotypic space 
with many local peaks separated by
fitness valleys. On such a rugged fitness landscape, a low fitness
population initially climbs a fitness peak until it encounters a local
peak where it gets trapped since a better peak lies some mutational
distance away. In a population of realistic size, it takes a finite
time for an adaptive mutation to arise and the fitness stays constant
during this time (stasis). Once some beneficial mutants become
available, the fitness increases quickly as the population moves to a
higher peak where it can again get stuck. Such dynamics alternating
between stasis and rapid changes in fitness go on until the population reaches
the global maximum.

\begin{figure}
\begin{center}
\includegraphics[angle=270,scale=0.4]{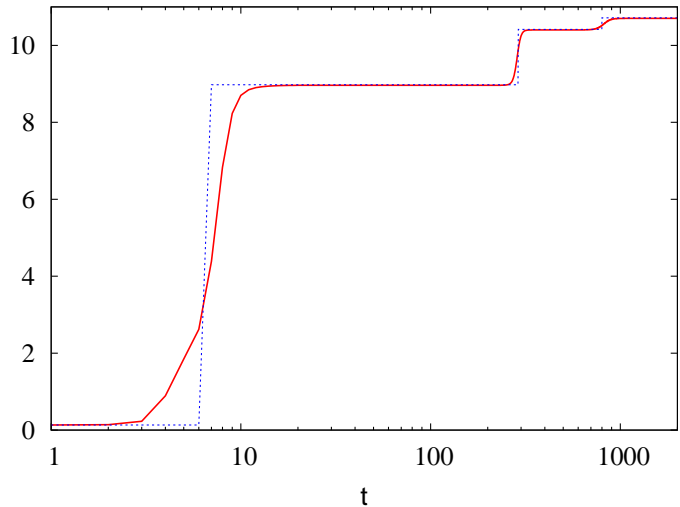}
\caption{(Color online)Punctuated change in the average population fitness
  (dotted line) and the fitness of the most populated genotype (solid line) for
  an infinite population evolving on a maximally 
rugged fitness landscape. Here genome length $L=15$ and mutation
probability $\mu=10^{-4}$.}
\label{avgQ}
\end{center}
\end{figure}

This punctuated behavior of fitness is also seen in deterministic models that
assume infinite population size. An example of such a step-like
pattern for average fitness is shown in Fig.~\ref{avgQ}. A neat and
unambiguous way of defining a step is by considering the
fitness of the most populated genotype also shown in Fig.~\ref{avgQ}. 
Since large but
finite populations evolve deterministically at short times
\cite{Jain:2007a}, it is
worthwhile to study the punctuated evolution in models with infinite
number of individuals. 
In this article, we will briefly describe some exact results concerning
the dynamics of an infinitely large population on rugged fitness landscapes
\cite{Jain:2005,Jain:2007c}. We will find that the mechanism producing
the step-like behavior is not due to ``valley crossing'' as in finite
populations but when a fitter
population ``overtakes'' the less fit one as described in the subsequent
sections. 
 
\section{Quasispecies model and its steady state}

We consider an infinitely large population reproducing 
asexually via the elementary processes of selection and mutation. Each  
individual in the population carries a binary
string $\sigma=\{\sigma_1,...,\sigma_L \}$ of length $L$ where
$\sigma_i=0$ or $1$. The $2^L$ sequences are arranged on the
multi-dimensional Hamming space. 
The information about the environment is encoded
in fitness landscape defined as a map from the sequence space into the
real numbers and is generated by associating a non-negative real number
$W(\sigma)$ to each sequence $\sigma$. 
Fitness landscapes can be simple possessing some symmetry properties
such as permutation invariance, or complex devoid of any such symmetries 
\cite{Gavrilets:2004,Jain:2007b}. Fitness functions with single peak are an
example of simple fitness landscapes while rugged landscapes with many
hills and valleys belong to the latter class. 

The average population fraction ${\cal X}(\sigma,t)$ with sequence
$\sigma$ at time $t$ follows mutation-selection dynamics described by
the following discrete time equation \cite{Eigen:1971,Jain:2007b}
\be
{\cal X}(\sigma,t+1)= \frac{\sum_{\sigma'} p_{\sigma \leftarrow \sigma'} 
W(\sigma') 
{\cal X}(\sigma',t)}{\sum_{\sigma'} W(\sigma') {\cal X}(\sigma',t)}~.
\label{quasi}
\ee
The last two factors in the numerator of the above equation give the
population fraction when  a sequence $\sigma'$ copies itself with
replication probability $W(\sigma')$ since fitness is defined as the
average number of offspring produced per generation. 
After the reproduction process, point mutations are introduced independently 
at each locus of the sequence $\sigma'$ with probability $\mu$ per generation. 
Thus, a sequence $\sigma$ is obtained via mutations in $\sigma'$ with 
probability 
\be
p_{\sigma \leftarrow \sigma'} = \mu^{d(\sigma,\sigma')} (1 - \mu)^{L-d(\sigma,\sigma')}
\label{mut}
\ee
where the Hamming distance $d(\sigma,\sigma')$ is the number of point 
mutations in which the sequences $\sigma$ and $\sigma'$ differ. 
The denominator of (\ref{quasi}) is the average fitness of the population at time $t$
which ensures that the density ${\cal X}(\sigma,t)$ is conserved. 

The stationary state of the quasispecies equation (\ref{quasi}) has been studied
extensively in the last two decades for various fitness
landscapes. These numerical and analytical studies
have shown that for most landscapes, there exists a critical mutation
rate $\mu_c$ below which the
population forms a quasispecies consisting of fittest genotype and its
closely related mutants while above it, the population delocalises
over the whole sequence space. This {\it error threshold} phenomenon can be
easily demonstrated for a single peak fitness landscape defined as 
\be
W(\sigma)=W_0
\delta_{\sigma,\sigma_0}+(1-\delta_{\sigma,\sigma_0})~,~W_0 > 1
\ee
where $\sigma_0$ is the fittest sequence. In the limit 
$\mu \to 0, L \to \infty$ keeping $U=\mu L$ fixed, the frequency of the fittest
sequence in the steady state of (\ref{quasi}) is given by
\be
{\cal X}(\sigma_0)= \frac{W_0 e^{-U}-1}{W_0-1}
\ee
which is an acceptable solution provided $U \leq U_c=\ln W_0$. For $U >
U_c$, selection is unable to counter the delocalising effects of
mutation and the population can not be maintained at the fitness
peak. 
For a discussion of error threshold phenomenon on other fitness
landscapes and generalisations of the basic quasispecies equation 
(\ref{quasi}), we refer the reader to \cite{Jain:2007b}.  
\section{Quasispecies dynamics on rugged fitness landscapes}

We now turn our attention to the dynamical evolution of ${\cal
  X}(\sigma,t)$ on  rugged fitness
landscapes. We consider maximally rugged fitness landscapes for which 
the fitness $W(\sigma)$ is a random variable 
chosen independently from a common distribution. 
It is useful to introduce the unnormalised population defined as 
\be
{\cal Z}(\sigma,t)={\cal X}(\sigma,t) \prod_{\tau=0}^{t-1} \sum_{\sigma'} W(\sigma') {\cal X}(\sigma',t)
\ee
in terms of which the nonlinear evolution 
(\ref{quasi}) reduces to the following linear iteration 
\be
{\cal Z}(\sigma,t+1)= \sum_{\sigma^{'}} p_{\sigma \leftarrow \sigma^{'}} 
W(\sigma') 
{\cal Z}(\sigma',t)~.
\l{linear}
\ee
Since at the beginning of the adaptation process the 
population finds itself at a low fitness genotype, 
we start with the initial condition 
${\cal X}(\sigma,0)={\cal Z}(\sigma,0)=\delta_{\sigma,\sigma^{(0)}}$ 
where $\sigma^{(0)}$ is a randomly chosen sequence. 
For mutation probability $\mu \to 0$,  after one 
iteration we have
\be
{\cal Z}(\sigma,1) \sim \mu^{d(\sigma,\sigma^{(0)})} W(\sigma^{(0)})~.
\l{one-gen}
\ee
Thus in an infinite population model, each sequence gets populated in one
generation obviating the need for ``valley crossing'' which is
required for finite populations. Although an exact solution of 
(\ref{linear}) for $t > 1$ is not available, it is possible to obtain several
asymptotically exact results concerning  the most populated genotype
using a simplified version of the quasispecies dynamics.  
Numerical simulations of \cite{Krug:2003} showed that 
dynamical properties involving the most populated genotype 
are well described by a simplified model which approximates the
population ${\cal Z}(\sigma,t)$ in (\ref{linear}) by 
\be
{\cal Z}(\sigma,t) \sim \mu^{d(\sigma,\sigma^{(0)})} W^t (\sigma)~,~t
> 1~.
\l{shell}
\ee
This model ignores mutations once each sequence has been populated and
 allows the population at each sequence to grow with its own fitness.  
However, a recent perturbative analysis in the small parameter $\mu$ 
shows that this approximation holds for highly fit sequences and at short times 
\cite{Jain:2007c}.

\begin{figure}
\begin{center}
\includegraphics[angle=270,scale=0.35]{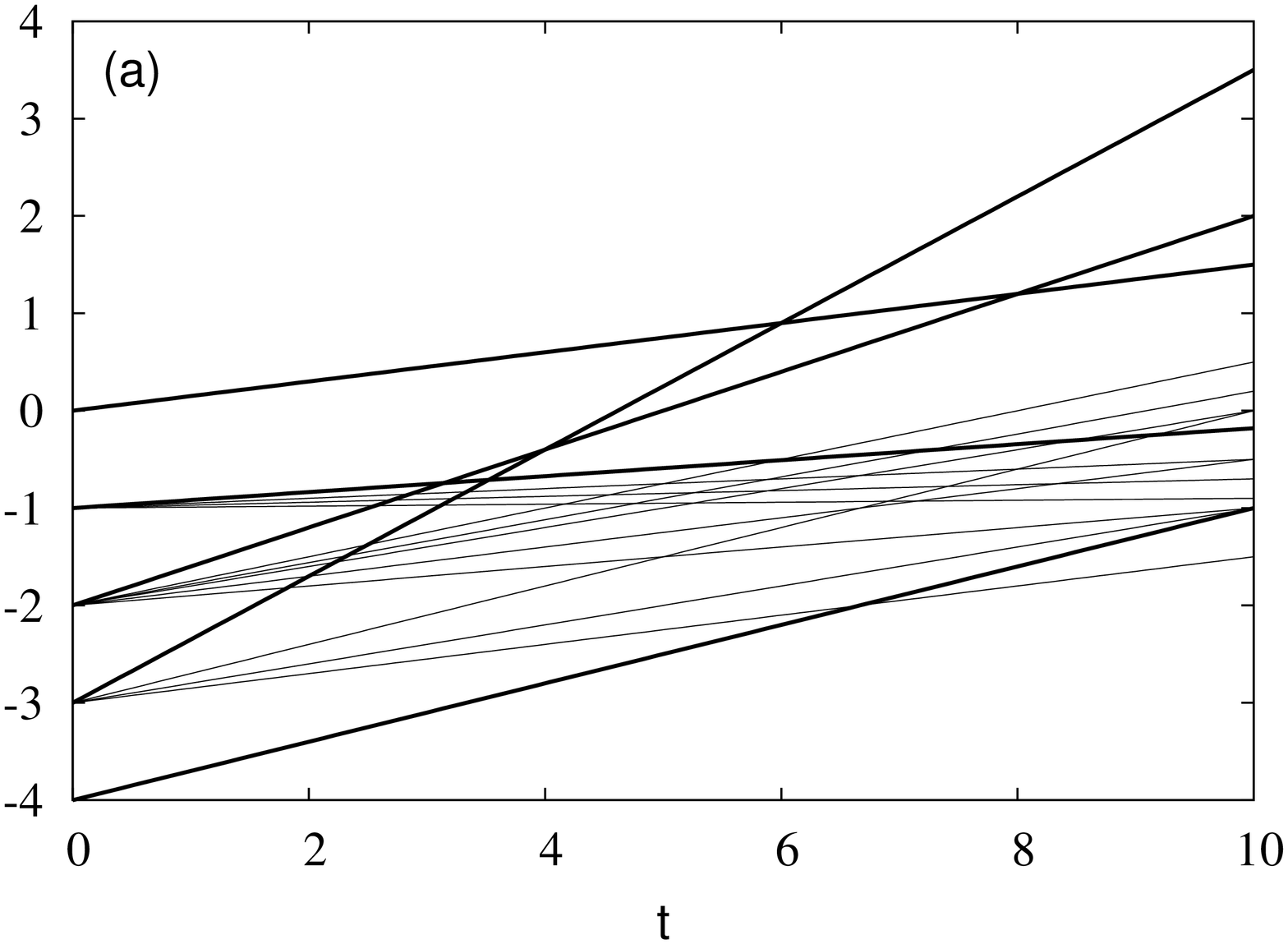}
\includegraphics[angle=270,scale=0.35]{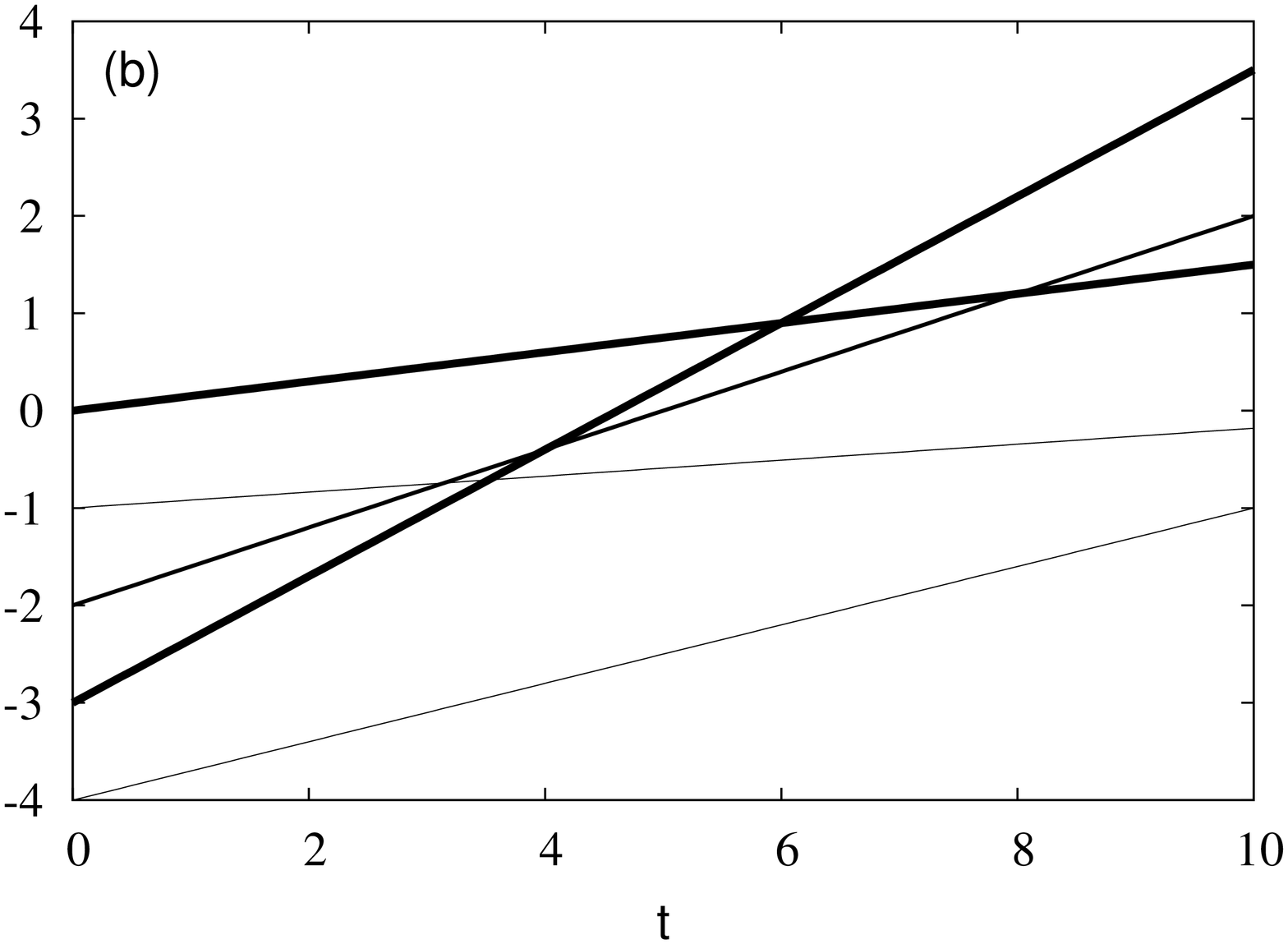}
\caption{(a) Evolutionary trajectories
  $E(\sigma,t)$ defined by (\ref{model}) for $L=4$. The bold lines
  have the largest fitness amongst the ${L \choose k}$ fitnesses at
  distance $k$ from the origin. (b) Evolutionary race: The sequence at
  distance $3$ is the most populated sequence (winner) while the one
  at distance $2$ is a record (contender).}
\label{talklines}
\end{center}
\end{figure}

Writing $W(\sigma)=e^{F(\sigma)}$ and rescaling time by $|\ln \mu|$ in
(\ref{shell}), we find that the
logarithmic population $E(\sigma,t)$ obeys the following linear
equation:
\be
E(\sigma,t)=-d(\sigma,\sigma^{(0)}) + F(\sigma)~t~.
\l{model}
\ee
The linear evolution of the (logarithmic) population of $2^L$ sequences for
$L=4$ is shown in Fig.~\ref{talklines}a. Since the initial population
fraction given by (\ref{one-gen}) is same for all the sequences
at constant Hamming distance $d(\sigma,\sigma^{(0)})$ from
$\sigma^{(0)}$, ${L \choose d}$ lines are seen to emanate
from the same intercept. However as the genotype with the largest
slope (fitness) 
at constant intercept has the potential to become the most
populated sequence, we arrive at the model in Fig.~\ref{talklines}b in
which $L+1$ genotypes are retained, each of whose fitness $F(k),
k=0,...,L$ is an independent but non-identically distributed variable 
\cite{Krug:2003,Jain:2005}. 

In a sequence $\{F(k)\}$ of random variables, a {\it record} is said to
occur at $m$ if $F(m) > F(k)$ for all $k < m$.
In Fig.~\ref{talklines}b, the sequences at
distance $k=0, 2$ and $3$ from the initial sequence are
records but the sequence at $k=2$ does not become a most populated
genotype. In order to qualify as a {\it jump}, it is not
sufficient to have a record fitness; the population should also be
able to overtake the current winner in minimum time. Due to the
overtaking time minimization constraint, the records and
jumps have different statistical properties which we describe briefly 
in the next subsections. 

\subsection{Statistics of records}

Although the record statistics for independent and identically
distributed (i.i.d.) random variables is well studied, much less is known
when the variables are not i.i.d.\cite{Nevzorov:2001}. Here we
have a situation in which $F(k)$ is a maximum of $\alpha_k={L \choose k }$
i.i.d. random variables. However, since
the $k$th record fitness $F(k)$ is the largest amongst $\sum_{j=0}^k
\alpha_j$ i.i.d. variables and there are $\alpha_k$ ways of choosing it, 
the probability $\tilde{P}_k$ that the $k$th fitness is a record is given by
\cite{Jain:2005,Krug:2005}
\be
\tilde{P}_k=\frac{{L \choose k}}{\sum_{j=0}^k {L \choose j}} 
\approx \frac{L-2k}{L-k}~,~k < L/2~.
\l{Ptildeshell}
\ee 
The meaning of the above distribution is intuitively clear: as it is
easier to break records in the beginning, the probability to find a
record is near unity for $k \ll L$ and it vanishes beyond $L/2$ because 
the global maximum typically occurs at this distance. 
The average 
number ${\cal{R}}$ of records can be obtained by simply 
integrating $\tilde{P}(k)$ over $k$ to yield ${\cal{R}} \approx (1-\ln
2)L$. It is also possible to find the typical spacing
$\tilde{\Delta}(j)$ between the $j$th and $(j+1)$th record where we have
labeled the last record (i.e. global maximum) as $j=1$. A
straightforward calculation shows that the typical inter-record
spacing falls as a power law given by \cite{Jain:2005} 
\be
\tilde{\Delta}(j) \approx  
\sqrt {\frac{L}{4 \pi j}} \;\;,\;\; j \gg 1 \;\;. \l{interrecshell}
\ee
The above expression indicates that the spacing between the last few 
records (i.e $j \sim {\cal{O}}(1)$) is of order $\sqrt{L}$, while most
of the records are crowded at the beginning which is consistent with
the behavior of the record occurrence probability
(\ref{Ptildeshell}). 

\subsection{Statistics of jumps}

The calculation of jump statistics \cite{Jain:2007c} 
is more involved than that of
records because a jump event requires a minimization of the overtaking
time. This constraint imposes a condition on the fitnesses of the
squences that can possibly overtake the current leader in a time
interval between $t$ and $t+dt$. The sequence at distance $k'$
can overtake the $k$th one (with fitness $F$) at time $t$ if the 
fitness $F(k')=(E(k,t)+k')/t$ and at time $t+dt$, $dt/t \to 0$ if 
\be
F(k')=\frac{E(k,t+dt)+k'}{t+dt}= F+\frac{k'-k}{t}-\frac{k'-k}{t^2} dt+{\cal O}(dt^2)~. \no
\ee
Then the total collision rate $W_{k',k}(F,t)$ with 
which the $k$th sequence is overtaken by the $k'$th one is given as 
\be
W_{k',k}(F,t) \approx \frac{k'-k}{t^2} ~p_{k'} \left(F+ \frac{k'-k}{t} \right)~,~k' > k~
\l{coll}
\ee
where $p_k(F)$ is the distribution of the maximum of $\alpha_k$
i.i.d. random variables distributed according to $p(F)$ with support
over the interval $\left[ F_\m,F_\M \right]$. Using this collision
rate, we can write
the probability ${\cal P}_{k',k}(t)$ that the sequence at distance 
$k'$ overtakes the $k$th one at time $t$ as
\be
{\cal P}_{k',k}(t)= \int_{F_\m}^{F_\M} dF~W_{k',k}(F,t)~P_{k}(F,t)
\label{basic}
\ee
where the probability $P_k(F,t)$ that the $k$th sequence has the
largest population at time $t$ is given by  
\be
P_{k}(F,t)=p_k(F)~ \prod_{\substack{j=0\\j \neq k}}^L \int_{F_\m}^{F+\frac{j-k}{t}} dF'~p_j(F')~. 
\l{max}
\ee
Note that unlike the records, the jump properties depend on the underlying
distribution of the random variables. Below we present some results
when the distribution $p(F)=e^{-F}$. 

Integrating (\ref{basic}) over time, the  probability distribution 
$P_{k',k}$ that $k$th sequence is overtaken by $k'$th sequence
is obtained, 
\be
P_{k',k} \approx \sqrt{\frac{L}{\pi k (L-k)}}~\left(\frac{k'-k}{2 k}\right)~e^{-\frac{L (k'-k)^2}{4k (L-k)}}~,~k< k' < L/2~.
\ee
This form of the distribution implies that the overtaking sequence $k'$ is 
located within ${\cal O}(\sqrt{k})$ distance of the overtaken sequence $k$. 
Thus the typical spacing between successive jumps 
for large $k$ is roughly constant and goes as $\sqrt{L}$ unlike in the
case of records discussed in the last subsection. 
The jump distribution $P_k$ for a jump to occur at distance $k$ 
is obtained by integrating over $k'$ and we have \cite{Jain:2007c}
\be
P_k \approx \sqrt{\frac{L}{\pi k (L-k)}} ~\theta_H \left(\frac{L}{2}-k \right)
\ee
where $\theta_H$ is the Heaviside step function which takes care of
the fact that the record
distribution (\ref{Ptildeshell}) vanishes at distance $L/2$. Instead
of integrating over time, 
by summing over the space variables $k, k'$ in (\ref{basic}), the
probability $P(t)$ that a
jump occurs at time $t$ can be obtained and is given by \cite{Jain:2007c}
\be 
P(t)=\sqrt{\frac{L}{4 \pi}}~\frac{1}{t^2} ~\mathrm{sech} \left(\frac{1}{2t} \right)~.
\l{Jt-shell}
\ee
The heavy tail distribution $P(t) \sim t^{-2}$ can be understood using
a simple argument \cite{Krug:2003} and implies that mean overtaking
time is infinite. Finally, by either summing $P_k$ over $k$ or
integrating $P(t)$ over time, the total number of jumps ${\cal J}$ are found to
be  $\sqrt{L \pi}/2$ which is much smaller than the number of records
${\cal R}$. 

\section{Summary}

In this article, we discussed the steady state and the dynamics of the
quasispecies model which describes a self-replicating population
evolving under mutation-selection dynamics. On rugged fitness
landscapes, the population fitness increases in a punctuated fashion
and we described several exact results concerning this mode of
evolution. 
Our recent simulations indicate that the $1/t^2$ law in (\ref{Jt-shell}) for
the deterministic populations also holds for finite stochastically
evolving populations \cite{Jain:2007c}. At present, we do not have an
analytical understanding of the latter result but it should be
possible to test this law in long-term experiments such as those of 
\cite{Elena:2003a} on {\it E. Coli}.

Acknowledgements: I am very grateful to Prof. J. Krug for introducing me to
the area of theoretical evolutionary biology. I also thank the
organisers of the Statphys conference at IIT, Guwahati for 
giving me an opportunity to present my work.



\begin{thebibliography}{47}
\expandafter\ifx\csname natexlab\endcsname\relax\def\natexlab#1{#1}\fi
\expandafter\ifx\csname bibnamefont\endcsname\relax
  \def\bibnamefont#1{#1}\fi
\expandafter\ifx\csname bibfnamefont\endcsname\relax
  \def\bibfnamefont#1{#1}\fi
\expandafter\ifx\csname citenamefont\endcsname\relax
  \def\citenamefont#1{#1}\fi
\expandafter\ifx\csname url\endcsname\relax
  \def\url#1{\texttt{#1}}\fi
\expandafter\ifx\csname urlprefix\endcsname\relax\def\urlprefix{URL }\fi
\providecommand{\bibinfo}[2]{#2}
\providecommand{\eprint}[2][]{\url{#2}}


\bibitem[{\citenamefont{Darwin}(1859)}]{Darwin:1859}
\bibinfo{author}{\bibfnamefont{C.}~\bibnamefont{Darwin}},
  \emph{\bibinfo{title}{The origin of species by means of natural selection}}
  (\bibinfo{publisher}{John Murray, London},
  \bibinfo{year}{1859}).

\bibitem[{\citenamefont{Novella et~al.}(1995)\citenamefont{Novella, Duarte,
  Elena, Moya, Domingo, and Holland}}]{Novella:1995}
\bibinfo{author}{\bibfnamefont{I.}~\bibnamefont{Novella}},
  \bibinfo{author}{\bibfnamefont{E.}~\bibnamefont{Duarte}},
  \bibinfo{author}{\bibfnamefont{S.}~\bibnamefont{Elena}},
  \bibinfo{author}{\bibfnamefont{A.}~\bibnamefont{Moya}},
  \bibinfo{author}{\bibfnamefont{E.}~\bibnamefont{Domingo}}, \bibnamefont{and}
  \bibinfo{author}{\bibfnamefont{J.}~\bibnamefont{Holland}},
  \bibinfo{journal}{Proc. Natl. Acad. Sci. USA} \textbf{\bibinfo{volume}{92}},
  \bibinfo{pages}{5841} (\bibinfo{year}{1995}).

\bibitem[{\citenamefont{Elena and Lenski}(2003)}]{Elena:2003a}
\bibinfo{author}{\bibfnamefont{S.~F.} \bibnamefont{Elena}} \bibnamefont{and}
  \bibinfo{author}{\bibfnamefont{R.~E.} \bibnamefont{Lenski}},
  \bibinfo{journal}{Nat. Rev. Genet.} \textbf{\bibinfo{volume}{4}},
  \bibinfo{pages}{457} (\bibinfo{year}{2003}).


\bibitem[{\citenamefont{Jain and Krug}(2007{\natexlab{a}})}]{Jain:2007a}
\bibinfo{author}{\bibfnamefont{K.}~\bibnamefont{Jain}} \bibnamefont{and}
  \bibinfo{author}{\bibfnamefont{J.}~\bibnamefont{Krug}},
  \bibinfo{journal}{Genetics} \textbf{\bibinfo{volume}{175}},
  \bibinfo{pages}{1275} (\bibinfo{year}{2007}{\natexlab{a}}).

\bibitem[{\citenamefont{Jain and Krug}(2005)}]{Jain:2005}
\bibinfo{author}{\bibfnamefont{K.}~\bibnamefont{Jain}} \bibnamefont{and}
  \bibinfo{author}{\bibfnamefont{J.}~\bibnamefont{Krug}}, \bibinfo{journal}{J.
  Stat. Mech.: Theor. Exp.} p. \bibinfo{pages}{P04008} (\bibinfo{year}{2005}).

\bibitem[{\citenamefont{Jain}(2007{\natexlab{a}})}]{Jain:2007c}
\bibinfo{author}{\bibfnamefont{K.}~\bibnamefont{Jain}},
  \bibinfo{journal}{Phys. Rev. E} \textbf{\bibinfo{volume}{76}},
  \bibinfo{pages}{031922} (\bibinfo{year}{2007}{\natexlab{a}}).

\bibitem[{\citenamefont{Gavrilets}(2004)}]{Gavrilets:2004}
\bibinfo{author}{\bibfnamefont{S.}~\bibnamefont{Gavrilets}},
  \emph{\bibinfo{title}{Fitness landscapes and the origin of species}}
  (\bibinfo{publisher}{Princeton University Press},
  \bibinfo{year}{2004}).

\bibitem[{\citenamefont{Jain and Krug}(2007{\natexlab{b}})}]{Jain:2007b}
\bibinfo{author}{\bibfnamefont{K.}~\bibnamefont{Jain}} \bibnamefont{and}
  \bibinfo{author}{\bibfnamefont{J.}~\bibnamefont{Krug}}, in
  \emph{\bibinfo{booktitle}{Structural approaches to sequence evolution:
  Molecules, networks and populations}}, edited by
  \bibinfo{editor}{\bibfnamefont{U.}~\bibnamefont{Bastolla}},
  \bibinfo{editor}{\bibfnamefont{M.}~\bibnamefont{Porto}},
  \bibinfo{editor}{\bibfnamefont{H.}~\bibnamefont{Roman}}, \bibnamefont{and}
  \bibinfo{editor}{\bibfnamefont{M.}~\bibnamefont{Vendruscolo}}
  (\bibinfo{publisher}{Springer, Berlin}, \bibinfo{year}{2007}{\natexlab{b}}),
  pp. \bibinfo{pages}{299--340}, \eprint{arXiv:q-bio.PE/0508008}.

\bibitem[{\citenamefont{Eigen}(1971)}]{Eigen:1971}
\bibinfo{author}{\bibfnamefont{M.}~\bibnamefont{Eigen}},
  \bibinfo{journal}{Naturwissenchaften} \textbf{\bibinfo{volume}{58}},
  \bibinfo{pages}{465} (\bibinfo{year}{1971}).


\bibitem[{\citenamefont{Krug and Karl}(2003)}]{Krug:2003}
\bibinfo{author}{\bibfnamefont{J.}~\bibnamefont{Krug}} \bibnamefont{and}
  \bibinfo{author}{\bibfnamefont{C.}~\bibnamefont{Karl}},
  \bibinfo{journal}{Physica A} \textbf{\bibinfo{volume}{318}},
  \bibinfo{pages}{137} (\bibinfo{year}{2003}).


\bibitem[{\citenamefont{Nevzorov}(2001)}]{Nevzorov:2001}
\bibinfo{author}{\bibfnamefont{V.B.}~\bibnamefont{Nevzorov}},
  \emph{\bibinfo{title}{Records: Mathematical Theory}}
  (\bibinfo{publisher}{Providence, RI: American Mathematical Society}, \bibinfo{year}{2001}).


\bibitem[{\citenamefont{Krug and Jain}(2005)}]{Krug:2005}
\bibinfo{author}{\bibfnamefont{J.}~\bibnamefont{Krug}} \bibnamefont{and}
  \bibinfo{author}{\bibfnamefont{K.}~\bibnamefont{Jain}},
  \bibinfo{journal}{Physica A} \textbf{\bibinfo{volume}{358}},
  \bibinfo{pages}{1} (\bibinfo{year}{2005}).





\end{thebibliography}
\end{document}